\documentclass[11pt]{article}
\usepackage{moriond,epsfig}

\def\PRL{\em Phys. Rev. Lett.}
\def\PRD{{\em Phys. Rev.} D}

\def\be{\begin{equation}}
\def\ee{\end{equation}}
\def\bea{\begin{eqnarray}}
\def\eea{\end{eqnarray}}

\begin{document}
\vspace*{4cm}
\title{QCD AND HADRONIC INTERACTIONS \\
EXPERIMENTAL SUMMARY OF MORIOND '03}

\author{ P.M\"attig }

\address{University of Wuppertal\\
Wuppertal, Germany}

\maketitle\abstracts{The broad progress in QCD studies during the
last years is summarised.}

\section{QCD evidence: The last 24 years}

Next year's Rencontre de Moriond can celebrate the 25$^{th}$
anniversary of the direct observation of the gluon. With this
discovery in $e^+e^-$ collisons at DESY \cite{bib-gluondiscovery}
in 1979 QCD became the accepted theory of strong interaction and
boosted the confidence in the Standard Model. Twenty - four years
later the status of QCD is excellent. The only free parameter
$\alpha _s$ is measured in a lot of observables and different
kinds of reactions yielding consistent values and an overall
precision of 2-3\% . As a function of $Q^2$ the strong coupling
exhibits the expected strong energy variation, the most apparent
evidence for quantum corrections in gauge theories. Also the gluon
self coupling has been directly observed and found to agree with
theory.

\begin{figure}
\hskip 5.5cm \psfig{figure=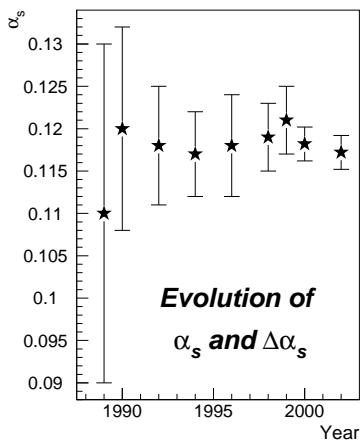,height=7cm} \caption{Evolution
of the value of $\alpha _s$ collected from various talks by
S.Bethke and the Particle data group \label{fig:alfaseva}}
\end{figure}

The progress in understanding QCD over the last two decades is
expressed in the shrinking uncertainty of $\alpha _s (Z^0)$ as
shown in Fig.~\ref{fig:alfaseva}. Whereas before LEP it was not
known to better than 20$\% $, during the last 10 years the
uncertainty was reduced by an order of magnitude.

Having praised the huge successes, let us turn to the basic
limitation of QCD tests: the beautifully displayed running of
$\alpha _s$, meaning a large value at low $Q^2$, causes it its
precision to be limited to 'some percent'. Theoretical
uncertainties and experimental signatures are far more complicated
than those for electroweak processes. The precision on $\alpha _s$
is some factor 200 below the one for the weak coupling, not to
speak of $\alpha _{em}$.

Both progress and difficulties are reflected in the main
directions in experimental QCD:

\begin{itemize}
\item[1.] Fix the one free parameter $\alpha _s(Z^0)$.
\item[2.] Develop models for long - distance contributions by invoking
additional symmetries or universal parameters motivated by QCD,
which have to be experimentally constrained.
\item[3.] Search for deviations from QCD as evidence for the New Physics.
\end{itemize}

Theoretical progress~\cite{bib-lynneorr} is substantial, however
experiments can easily match it. Since the discovery of gluon jets
an astounding experimental progress provides completely new ways
to test QCD: a huge extension of the kinematic range, highly dense
QCD matter, an increase of luminosity by several orders of
magnitude and high precision detectors to study special (heavy)
flavours. All this progress was reflected at this conference. Here
we will summarize the experimental contributions at this
conference starting with the highest energy scales and then
turning to lower and lower ones, before discussing searches for
New Physics.


\section{The value of $\alpha _s$}

One of the most exciting features of QCD is the relatively rapid
variation of its coupling with the energy scale. It is intimately
related to confinement and asymptotic freedom that are so unique
to QCD. Whereas up to recently $\alpha _s$ has been determined in
individual experiments at only one particular value of $Q^2$,
recent experiments span a large enough energy range with
sufficient precision to observe this running. At this Rencontre
such measurements were reported from HERA and LEP. Measurements
are based on jet - rates at different
$E_T^{jet}$~\cite{bib-vanmechelen} and $F_2$~\cite{bib-werner} at
HERA , or event shape analyses at different c.m. energies at
$e^+e^-$ colliders, either from LEP alone \cite{bib-rudolph} or
combined with resurrected JADE data \cite{bib-kluth}.

\begin{table}[t]
\caption{Values of $\alpha _s$ from measurements of energy
variations of QCD observables. In case of the $F_2$ results, the
theory error includes the model dependence. The first error on
$F_2$ from H1 includes both statistical and systematic effects.
\label{tab:alphas}} \vspace{0.4cm}
\begin{center}
\begin{tabular}{|c|c|c|}
\hline
& &  \\
  Method & ref & $\alpha _s(M_Z)$ (errors: stat, syst, theo) \\
  \hline
  Jet rates HERA photoproduction & \cite{bib-vanmechelen} &
       0.1212$\pm$ 0.0017 $^{+0.0023}_{-0.0031}$ $^{+0.0028}_{-0.0027}$
       \\ \hline
  Jet rates LEP & \cite{bib-rudolph} &
       0.1196$\pm$ 0.0017 $\pm $ 0.0010 $\pm $ 0.0049
       \\ \hline
  $F_2$ HERA (H1) & \cite{bib-werner} &
       0.1150$\pm$ 0.0017  $ \pm 0.0051$\\
       \hline
$F_2$ HERA (ZEUS) & \cite{bib-werner} &
       0.1166$\pm$ 0.0008  $\pm 0.0048$ $\pm 0.0053$ \\
        \hline

\end{tabular}
\end{center}
\end{table}

These measurements display beautifully the running of $\alpha _s$.
Assuming its QCD evolution with $Q^2$, they can be combined to
yield $\alpha _s(M_Z)$ as listed in table~\ref{tab:alphas}. These
values are in excellent agreement with each other, although they
use very different procedures! In addition they coincide with
individual measurements at single energies.

One notable aspect of these results is the dominance of the
theoretical uncertainty over the experimental ones. This
emphasizes the need for a better theoretical understanding. Since
one tries to combine $\alpha _s(M_Z)$ from various methods and
experiments to obtain the 'world average', it also points to the
need for commonly accepted and reliable procedures to estimate
values and uncertainties.


\section{A word on QCD uncertainties}

Estimating theoretical uncertainties means extrapolation into
something unknown - evidently a delicate effort. However, without
proper estimation of those uncertainties QCD tests will be
virtually impossible. A deviation would always be attributed to
the lack of higher order QCD calculations.

Current QCD uncertainties are par convention estimated by varying
the QCD scale between [0.5,2]. That this is not always sufficient
has been shown at this conference for calculations of the Higgs
production cross section and of $\gamma $ production at the LHC
\cite{bib-harlander,bib-schmidt}. For both processes uncertainties
estimated in lower order calculations were grossly underestimated.
These discrepancies may just be accidental, however, they may
point to the need of performing a systematic study of LO vs NLO vs
NNLO calculations to find ways of estimating theoretical
uncertainties.

Having said this, it should be clear that whatever procedure one
defines, theoretical uncertainties will not have a well defined
probability assigned as the experimental uncertainty.

It as an encouraging initiative that theorists and
experimentalists within the LEP - QCD group
\cite{bib-rudolph,bib-salamprivate} collaborate to define a
procedure for consistently estimating the theoretical
uncertainties of QCD measurements. Maybe it should be extended to
other QCD processes as well.


\section{Heavy Quark Production}

The production of heavy quarks is a beautiful test ground of QCD
for experimental and theoretical reasons

\begin{itemize}
\item It allows rather firm theoretical predictions since their
masses are large compared to the QCD scale $\Lambda $, rendering
perturbative expansions rather safe.

\item Because heavy quarks are suppressed in the
fragmentation process, the measurement of a charmed or bottom
hadron in a jet can almost unambiguously be associated to the
production of charm and bottom quarks at a hard scale.
\end{itemize}

The additional interest in heavy quarks is that in many extensions
of the Standard Model they are a harbinger of New Physics. Charm,
even more so bottom physics, and in the future top quarks will be
focal points of current and future collider experiments.

The lightest of these heavy quarks is the charm quark. And indeed,
cross section measurements from $\gamma \gamma $ interactions and
ep - collisions are in agreement with the theoretical NLO
predictions \cite{bib-boehrer,bib-bertolin}, which, however, still
have sizeable uncertainties of $\sim $ 30$\% $. More detailed
studies allow the discrimination between models. For example, in a
recent measurement ZEUS found that in resolved photoproduction
charm jets tend to be aligned with the photon direction as
expected from QCD calculations.

Agreement with the much safer theoretical expectations of $\sim $
5 $\% $ uncertainty is also found for pair production of top
quarks \cite{bib-cabrera}. Such measurements can only be performed
in $p\bar{p}$ collisions at the Tevatron. Run I experiments at
$\sqrt {s}$ = 1.8 TeV yielded a precision of some 25$\% $. For Run
II at $\sqrt {s}$ = 1.96 TeV the cross section is expected to
increase by about 30$\% $. For the first time top quarks have been
observed at these high energies. Within the limited statistics the
cross section measurements by CDF and D0 agree with the
prediction. With higher luminosity these will become an
interesting QCD test.

Whereas data and theory coincide for the lightest and the heaviest
of the heavy quarks, measurements of the cross section for bottom
quarks appear to be higher than QCD predictions. This is known
since several years and disagreements of some 3-4 standard
deviations between QCD NLO predictions and experiments have again
been shown at this conference. The measurements are performed in
the different parton environments of $p\bar{p}$, $\gamma \gamma $,
and $ep$ interactions \cite{bib-kajfasz,bib-boehrer,bib-klimek},
in different kinematical regions, and applying different methods
of bottom tagging and are therefore difficult to compare in
detail. Comparing instead the ratios of observed and expected
yields, one finds that almost in all circumstances three to four
times more bottom quarks are produced than expected. As a side
remark, the J/$\psi $ yields of HERA-B is in good agreement with
the theoretical expectation \cite{bib-conde}.

The consistently higher measurements are stunning. However,
theoretical evaluations~\cite{bib-binneweiser,bib-nason} performed
a couple of years ago show that at least the CDF data are
consistent with theory, if QCD effects are more carefully
included, such as resummed NLL calculation merged with a NLO fixed
order calculation and the non-perturbative part of the bottom
fragmentation function. Accounting for these, reduces the ratio
data over from 2.9 to 1.7 which can be accommodated by theoretical
uncertainties.

\begin{figure}
\hskip 6.5cm \psfig{figure=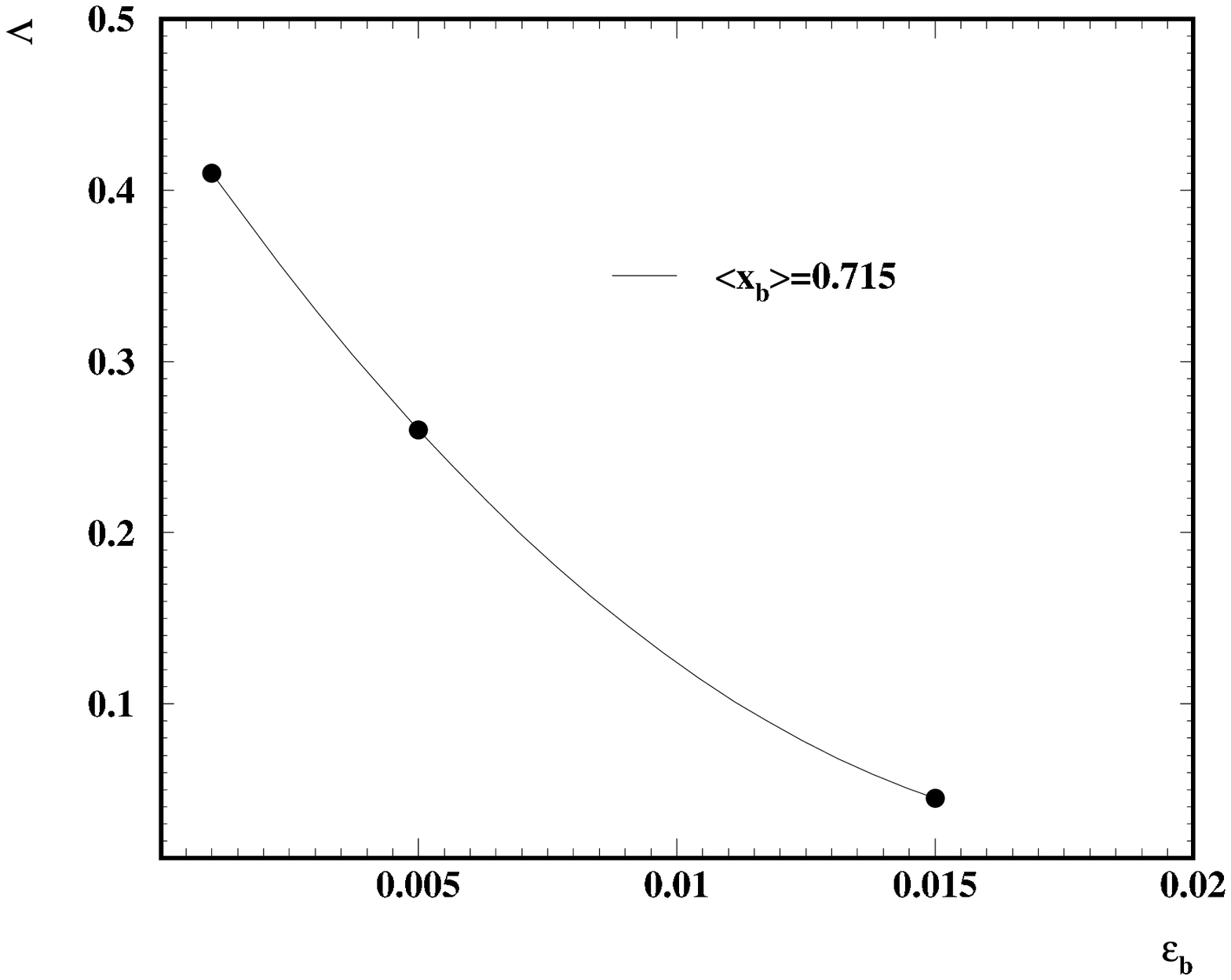,height=5cm}
\hskip 14cm\psfig{figure=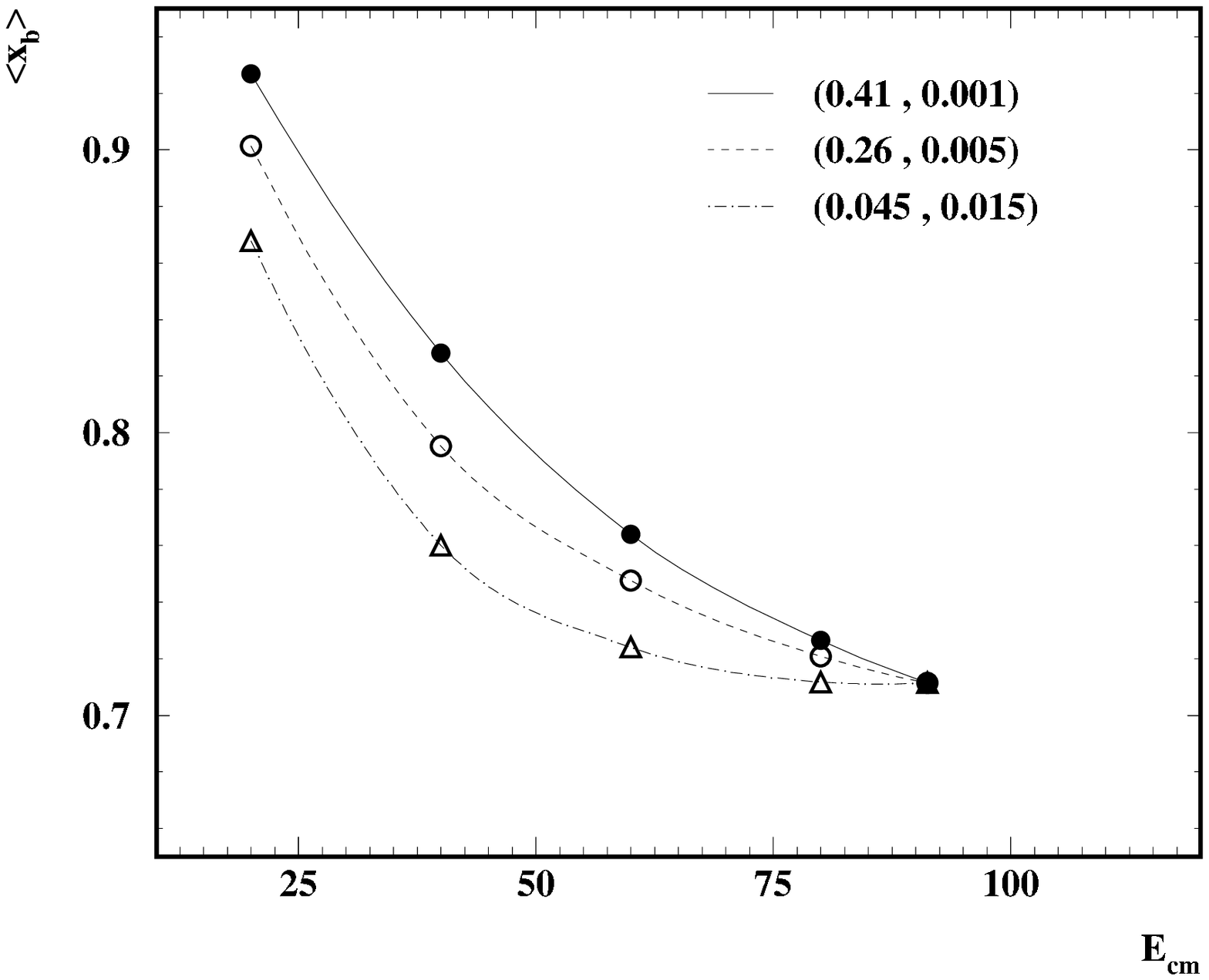,height=5cm}
\caption{a.Correlation of the parameters $\epsilon _b$ and
$\Lambda _{QCD}$ leading to $<x_b>$ = 0.715, b. $<x_b>$ as a
function of the c.m. energy for various combinations of $\Lambda
_{QCD}$ and $\epsilon _b$ \label{fig:epsb-lamb}}
\end{figure}

A word on the non-perturbative part. The best measurements of
bottom hadronisation have been obtained in $Z^0$ decays at LEP and
SLC. These measurements are usually parametrised within PYTHIA
using the Petersen et al. fragmentation function with just one
free parameter $\epsilon _b$. This $\epsilon _b$ is then assumed
for other kinds of collision. However, at this stage one has to be
careful: $\epsilon _b$ is not a fundamental parameter but is
specific to a certain version and parameter set of PYTHIA. This
can be seen in Fig.~\ref{fig:epsb-lamb}a where the average scaled
energy $<x>$ = 0.714 measured at LEP and SLC is reproduced by
combinations of the QCD scale parameter $\Lambda _{PYTHIA}$ and
$\epsilon _b$. For extracting the bottom cross section in
$p\bar{p}$, $\gamma \gamma $, and $ep$ interactions, one has to
integrate over a range of hard scales different from $M_Z$. The
bottom fragmentation function at these energies have to be evolved
and therefore depend strongly on the QCD scale $\Lambda $ as shown
in Fig.~\ref{fig:epsb-lamb}b. Such different fragmentation
functions may lead to quite different acceptance corrections and
thus different apparent cross sections when bottom hadrons are
selected via lepton energies or decay lengths. Therefore the LEP
results have to be applied with great care.

In an alternative procedure, following theoretical suggestions the
non-pertur\-bative hadronisation has been unfolded from the energy
distribution of bottom hadrons in $e^+e^- \rightarrow Z^0$ at LEP
and SLC in a model independent way~\cite{bib-BenHaim}. Using the
NLO calculation for bottom quarks, the de-convolution of the
hadronisation into bottom hadrons was derived by developing the
observed energy distribution of bottom hadrons into moments. The
results disfavour the commonly used Petersen etal. fragmentation
function but show that the Bowler and Lund parametrisations fit
best.

After the critical theoretical
comments~\cite{bib-binneweiser,bib-nason} it appears appropriate
to reanalyse the data such that the hard corrections and
non-perturbative effects are properly taken into account. It is a
rather unfortunate situation that experiments maintain to show
results exhibiting significant inconsistencies with theory without
attempting a reanalysis.


\section{News from inside the proton}

Decreasing in $Q^2$, there are new precise data on the strange
quark structure function \cite{bib-tzanov}. Vector meson
production in electro- and photoproduction are in good agreement
with perturbative QCD~\cite{bib-voss,bib-brown}. New results were
presented confirming the cross section dependence on the
transverse spin of the incoming proton \cite{bib-surrow}. This
interesting observation has no obvious QCD explanation, but models
with special assumptions can accommodate the data

A lot of interest in diffractive processes has been stirred after
the surprising discovery at HERA that a sizeable fraction of
events are produced by the exchange of a colour neutral object,
traditionally deemed pomeron. Since then a much more detailed
insight into diffractive physics both at HERA and the Tevatron has
been obtained indicating a large gluonic component in the
exchanged pomeron and allowing one to extract diffractive parton
densities~\cite{bib-vanmechelen2} in inclusive diffractive deep -
inelastic scattering. They can be consistently applied to dijet
and open charm production showing that factorisation holds within
HERA. However, transfering them to diffractive processes in
$p\bar{p}$ collisions in a straight forward manner does not work.
Double diffractive processes have been measured at the Tevatron
Run II~\cite{bib-gallinaro}. These processes are interesting in
view of ideas on the production of Higgs bosons through these
processes~\cite{bib-pechanski}. Ideally the very good mass
resolution from the scattered protons leads to a rather narrow
Higgs signal. However, a crucial question is, if further particles
are scattered into the detector under small angles which might
blur the signal.


\section{How colour flows}

One of the fundamental assumptions in experimental studies of QCD
is the correspondence between the observable hadrons and the
underlying partons, the concept of Local Parton-Hadron
Duality~\cite{bib-LPHD}. Several studies indicate that this is
true even down to very low $Q^2$. The $\ln (1/x) $ distribution in
$e^+e^-$ events reflects very nicely the expectation both at an
individual energy and for the c.m. energy
dependence~\cite{bib-kluth,bib-rudolph}. Also a more detailed look
into the fraction of particles perpendicular to the event plane of
three jet events accords with the expectation~\cite{bib-BenHaim}.

Recently studies have been performed on the non - leading colour
flow in events with four partons. These studies have been driven
by the observation of W - pairs at LEP. When both W - bosons decay
hadronically, colour may flow not only between the quarks from one
W, but there is the less likely flow between partons from
different W - bosons. Its probability cannot be calculated from
first principles, but rather models have to be invoked. Based on
LPHD, the favoured method at LEP is to study, if the hadron flow
between quarks from different W - bosons is enhanced. The
measurements are inconclusive as to the existence of colour
reconnection, however, allow the rejection of some
models~\cite{bib-straessner}.

Colour can also be reconnected in $Z^0$ decays with two hard
quarks and two hard gluons. To enrich those,events with rapidity
gaps in jets are selected and compared to
models~\cite{bib-giunta}. The experimental key issue is to
distinguish quark from gluon jets, which can, for example, be
achieved by identifying bottom quarks.

\begin{table}[t]
\caption{Qualification of models of colour reconnection using four
- jet events at the $Z^0$ and in W - pair production events at
LEP. For the LEP combined W - pair measurement the difference in
standard deviations of the observed flow to the model is also
listed. \label{tab:colour-recon}} \vspace{0.4cm}
\begin{center}
\begin{tabular}{|c|c|c|}
\hline
& &  \\
  Model  & at $Z^0$ & in W -pair events \\ \hline
ARIADNE  & ruled out &  agreement (-2.1$\sigma $) \\
RATHSMAN & ruled out &  not considered \\
HERWIG   & ok        &  agreement (-2.6$\sigma $) \\
SJOSTRAND-KHOZE I   & not considered & ruled out (-5.2$\sigma $) \\
                 \hline

\end{tabular}
\end{center}
\end{table}

Comparisons to models at the $Z^0$ and in W - pairs are listed in
Table~\ref{tab:colour-recon}. The different environments at the
$Z^0$ pole and in W - pair production also lead to different
consistencies with data, for example for ARIADNE. It means (not
unexpectedly) that in the different environments colour is
reordered differently. This complicates any extrapolation of
results obtained in $Z^0$ decays to W - pairs. Otherwise it would
help estimating possible distortions in reconstructing the W -
mass from fully hadronic decays at LEP.


\section{Hadronization }

The large amount of data from $Z^0$ decays at LEP and its
cleanliness will for a long time be the outstanding source of
information on how and which hadrons are formed inside a jet. At
this conference new data from $\gamma \gamma $ interactions
confirming the diquark model \cite{bib-lin} were presented.

An insight into the space - time picture of hadronisation can be
obtained by Bose - Einstein correlations. At LEP a lot of
measurements have been performed mainly using $\pi ^{\pm}\pi
^{\pm}$ pairs. Recently L3 and OPAL have observed Bose Einstein
correlations in $\pi^0$ pairs~\cite{bib-metzger}. At face value
the correlation length and strength measured by OPAL exceed those
of L3. However, it may just reflect the different kinematical
selections and thus a dependence of the correlation on the
specific kinematics. It is of particular interest, if Bose -
Einstein correlations also exist between particles coming from
different $W$ bosons pairs. Because of the potential distortion of
mass measurements of the $W$ boson, substantial effort has gone
into searching for these correlations. Whereas three of the LEP
experiments see no effect, DELPHI claims an enhanced correlation
of a pair of equally charged pions of similar
momentum~\cite{bib-vanremortel}.

\begin{figure}
\hskip 6cm
\psfig{figure=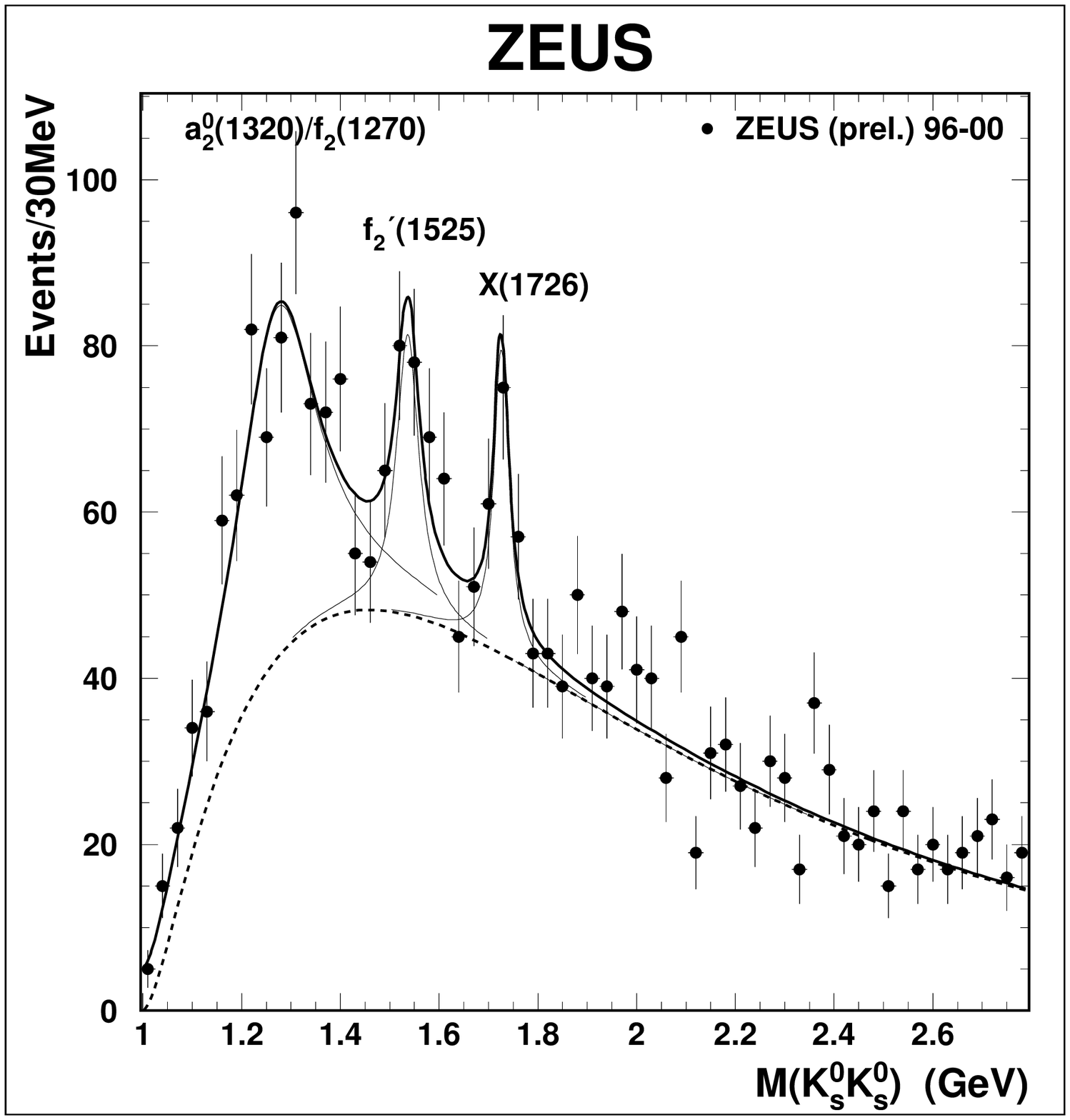,height=5cm}
\hskip 12cm \psfig{figure=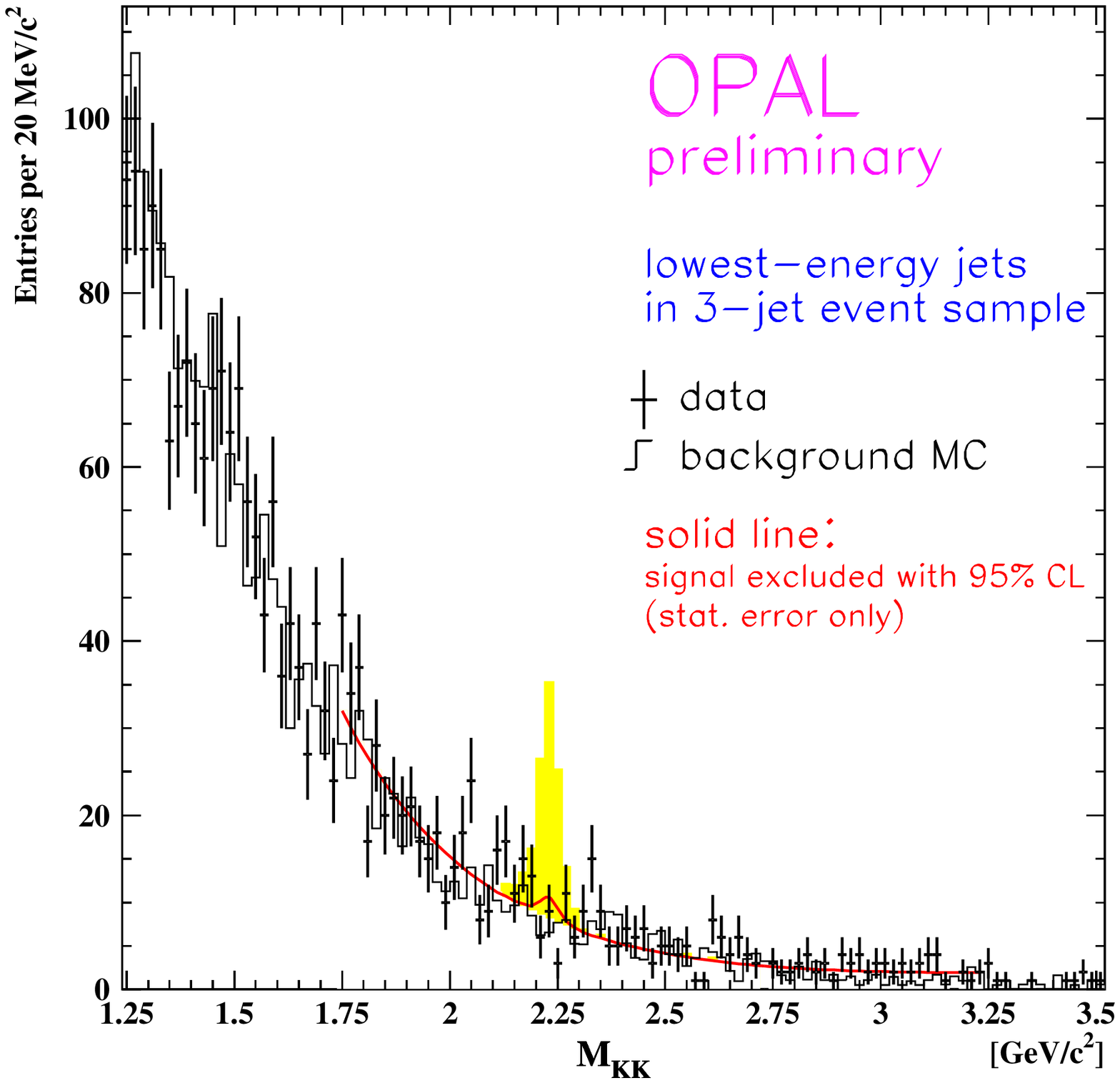,height=5cm}
\caption{$K^0_sK^0_s $ mass spectra. a. from photon-gluon fusion
at ZEUS, and b. in gluon jets at OPAL (the simulated signal around
2.25 GeV shows the sensitivity to an excess reported earlier by
L3). \label{fig:k0sk0s}}
\end{figure}

The existence of glueballs is a fundamental prediction of QCD, for
which, however,one has found at most indirect evidence. There are
suggestions that, apart from dedicated spectroscopic experiments,
glueballs may show up in processes with hard gluons. Since the
gluon self coupling is significantly stronger than its coupling to
quarks, the hope is that some of the many gluons produced coalesce
into a glueball. Such a hard gluon can be a gluon jet emitted in
$e^+e^-$ interactions or a virtual gluon in photon - gluon fusion
in ep collisions. ZEUS has analysed the mass spectrum of the
$K^0_sK^0_s$ system in such events~\cite{bib-paganis}. As can be
seen in Fig.~\ref{fig:k0sk0s}a, in addition to the known $a_2$ and
$f'_2$ they point to an enhancement at a mass of 1726 MeV where no
standard hadron exists. With their preferred background
parametrisation the $\sigma \times $Branching ratio for these
three resonances appear to be fairly similar.

By crossing the t - channel photon - gluon fusion of ep collision
one obtains the s - channel diagram for $e^+e^-$ annihilation into
a $qqg$ event. The $K^0_sK^0_s$ mass spectrum has been studied in
gluon jets by L3 and OPAL~\cite{bib-lepglueballs} at LEP. Neither
of them observes a peak around 1720 MeV (Fig.~\ref{fig:k0sk0s}b).

This does not necessarily disprove the ZEUS enhancement, since the
processes may, who knows, imply a different sensitivity to
glueballs. No doubt, a confirmation of the X(1726) in an gluonic
environment would be extremely interesting! Still, a justification
of the shape of the background with data instead from Monte Carlo
would be more convincing. For example, one might test if the
$K^+K^-$ shape can be reproduced by the simulation or if yields
for $f_2'$ production in the $\pi \pi $ and $K^0_sK^0_s$ decay
modes are consistent.


\section{Hadron decays}

The energy scale of hadronisation and hadron decays is the same.
However, in hadron decays the system starts from two (or three)
well defined quarks at a low energy scale. In these decays soft
gluons play an essential role. Since these can (as yet) not be
calculated from first principles, theorists invoke models based on
QCD and additional symmetries. The heavier the participating quark
is, the more reliable the model is. For the perturbative treatment
the essential requirement is $\Lambda /m_Q \ll 1$. In fact the
heaviest top quark does not need a model at all, since it decays
before forming hadrons. For the bottom and somewhat less for the
charm quark the Heavy Quark Effective Theory has provided a rather
precise understanding of decay properties. Alternatively, for the
up, down and strange quarks, each having a mass less than the QCD
scale $\Lambda _{QCD}$, models based on chiral symmetries have
been developed.

This substantial theoretical progress is accompanied by a stunning
experimental progress. The experimental precision on decay
properties of bottom and charm quarks, but also on the light
hadrons is extraordinary. Branching ratios of 10$^{-6}$ are now
routinely measured. I am aware that I do all these measurements
utterly injustice in just skimming them. In general these
measurements are in very good agreement with the expectations.
However the new level of precision also poses new and very
detailed challenges to the theoretical understanding.

Results on the pionium lifetime~\cite{bib-gotta} and the $\pi $N
scattering length~\cite{bib-schuetz} restrict the parameters of
chiral perturbation theory. Measurements on kaonic
atoms~\cite{bib-pgianotti}, respectively rare $K^0_s$
decays~\cite{bib-cheskov} were reported.

Several dedicated experiments, like FOCUS \cite{bib-kim}, CLEO
\cite{bib-dubrovin}, BABAR \cite{bib-meadows,bib-wagner}, E-835
\cite{bib-pastrone}, or multi - purpose experiments like CDF and
D0 \cite{bib-korn} collected a huge number of charmed events
pinning down charmonium transitions, CKM suppressed and rare charm
decays.

Of course, one of the main directions in today's high energy
physics is the understanding of CP - violation in bottom decays.
The huge data sets collected with BABAR and BELLE provide a wide
range of measurements. None of these clearly deviate from the
expectation, although some results, if confirmed with higher
statistics, may indicate that we fail to understand some aspects.
The study of CP violation is complemented by a broad program of
exploring the bottom hadrons. BABAR, BELLE and CLEO improve the
precision of the CKM matrix \cite{bib-warburton,bib-limosani}, and
determine tiny branching fractions of Bottom mesons
\cite{bib-vuagnin,bib-pakhlov,bib-wang,bib-dubosq,bib-pivk,bib-kuzmin,bib-coan},
CDF and D0 show the way towards measuring $B_0^S$ mixing
\cite{bib-vacavent}. This is really impressive progress!


\section{Heavy Ion Collisions}

Up to now we were going down in $Q^2$. It is unclear which $Q^2$
determines the measurements on Heavy Ion collisions presented at
this conference. Experiments at the SPS are continuing to evaluate
their data for the many potential signatures of the quark - gluon
plasma like strange particle production \cite{bib-vanleeuwen},
J/$\psi $ suppression \cite{bib-santos}, hyperon
production~\cite{bib-hibruno}, fluctuations \cite{bib-sako}, charm
production with new detectors \cite{bib-usai}.

Whereas some of these are consistent with the quark - gluon
plasma, others are not. The search for a deeper understanding of
Heavy Ion collisions has now turned to the higher energies of the
Relativistic Heavy Ion Collider RHIC. The focus of RHIC at this
Rencontre were on the existence and structure of jets. In brief,
the main observations at the RHIC experimenters
are~\cite{bib-velkovsa,bib-jacobs}
\begin{itemize}
\item[a.] Jets have been seen in $pp$ collisions at
$\sqrt {s}$ \ = \ 200 GeV.
\item[b.] Jet production is suppressed in central $Au-Au$
collisions at $\sqrt {s_{NN}}$ \ = \ 200 GeV (deemed 'jet
quenching').
\item[c.] For central collisions and $p_T \sim $ 2-4 GeV mesons are
produced as often as baryons, and baryons only a factor $\sim $
1.3 more often as antibaryons~\cite{bib-hippolyte}.
\end{itemize}

These observations are very intriguing: jet production would mean
a partonic interaction, whereas jet quenching in the dense medium
of central high energy nuclei collision might point to energy
losses of the partons due to gluon interaction. Thus at RHIC
densities partons, instead of nuclei would interact.

At this stage a few remarks from the perspective of a high energy
physicist. First of all, the jets discussed at RHIC are identified
by a leading particle with a $p_T$  of some 4 GeV. I.e. the jet
energy is only marginally above those energies where jets were
observed for the first time by evolved statistical arguments in
the very clean $e^+e^-$ environment by Mark II~\cite{bib-markII}.
They are substantially less energetic than the clear jets at
higher energy $e^+e^-$ collisions or $p\bar{p}$ collisions.

The evidence for jets in pp collisions resides on

\begin{itemize}
\item same side correlations showing an enhanced production
of particles of 2 - 4 GeV at a distance $\Delta \phi \sim $ 0.2 in
azimuth angle from the trigger particle,
\item enhanced particle activity at $\phi _{trigger} - \pi $,
\item charge correlations between the trigger particle and the high
$p_T$ particle around the trigger both of opposite and
(significantly less pronounced) same charge.
\cite{bib-jacobsprivate}.
\end{itemize}

Since 35 years we do know that jets will be produced in $NN$
collisions of 200 GeV. However, they will exist over a high
pedestal of minimum bias events. On this basis, are the
observations at RHIC sufficient to prove that what they see are
jets? All the above observations are consistent with the existence
of jets, however, they can also be due to simple resonance decays.
The parameters of the same side correlations point to two -
particle masses of $M^2\sim 2\cdot 4 \cdot (1-\cos 0.2)$, i.e. $M
\sim 0.5 $ GeV, typical of resonances. Resonances could also
explain opposite sign charge correlations of the leading
particles. The correlation at opposite $\phi $ may well reflect a
trigger bias.

A key question is then how much of the observed effect is due to
jets, respectively resonance decays. Assuming the extreme that all
the effect are due to resonances, then also the claim of jet -
quenching in Au-Au would have to be reconsidered. It needs more
detailed studies to really establish that a significant portion of
the observed correlations are due to jets. To exclude that the
observed structure just reflects some trigger bias or statistical
correlations, it would be reassuring if the data are compared with
a statistical distribution of resonances {\it before} decays.

Accepting the existence of jets at RHIC, those observed in $Au -
Au$ collisions show some unexpected properties. Whereas for
peripheral collisions the proton/pion ratio of $\sim $0.15 is as
small as in $e^+e^-$ jets, this ratio is $\sim 1$ for jets in
central collisions, at least for $p_T$ between $\sim $ 2 and 4
GeV. This may be even explainable by the abundance of protons in
the initial state. However, also the yield of antiprotons is about
the same as the one for $\pi ^-$ in central collisions for $p_T\
\sim $ 2 - 4 GeV. This is in sharp contrast to high energy jets in
other environments and calls for new ideas on the physics in high
energy collisions.

Several contributions~\cite{bib-interpret} at this Rencontre
provided additional experimental facts and attempts of an
interpretation. The outstanding question is, if the observed
structure arises from hard interactions at the quark - gluon level
and not from purely hadronic interactions, or from a mixture of
both. As pointed out at this conference, the planned measurements
on d-Au collisions will provide additional input, but then in
particular those extending to higher $E_T$.


\section{QCD and Beyond the Standard Model Physics}

The Standard Model is almost unchallenged, with the possible
exception that the finite $\nu $ masses may already be outside its
limits. Anyway, the Standard Model has conceptional deficiencies:
the proliferation of fermion generations and the similarity of its
interactions, the naturalness problems etc. call for a larger
theory. Now, that the Standard Model has been so beautifully
confirmed in the 100 GeV range, the focus of High Energy Physics
shifts more and more to the search for new effects. What can QCD
measurements reveal about New Physics?

In general the sensitivity of QCD processes is less than those
from the electroweak measurements due to the inherent theoretical
uncertainties in QCD predictions, as becomes evident from
comparing the precision on $\alpha _s$ to the one on $G_F$ or
$\alpha _{em}$, and (not unrelated) to the worse experimental
resolutions, apparent by comparing jet to electron or photon
resolutions.

The caution in interpreting deviations in QCD distributions as
evidence for new physics can be seen, for example, in the
distribution of dijet masses from the Tevatron~\cite{bib-hays}.
Uncertainties in the parton distribution functions, jet energy
resolutions etc. dilute the significance of any possible effects.
The excitement of the mid 90s about an apparent deviation of high
$E_T$ jet production, which in the end could well be explained by
alternative parton distribution functions, is still a warning. At
this conference we were reminded of uncertainties due to the
insufficient knowledge of the pdfs in view of potential signals of
small extra space dimensions~\cite{bib-ferrag} at the LHC.

Still, in some cases purely hadronic processes do constrain
theories beyond the Standard Model. LEP results on colour factors
from four - jet rates~\cite{bib-rudolph} allow one to exclude a
light gluino as anticipated for some SUSY parameters. Another
possibility is the search for bumps in the dijet mass. Whatever
the pdfs really are, whatever the energy resolution, in all
circumstances the mass distribution should be smooth. Measurements
by CDF and D0 does not show any such enhancement~\cite{bib-begel}
and can be translated into limits on models like excited quarks or
technicolour particles~\cite{bib-hays,bib-gallinaro}.


\section{The other part of the Standard Model}

Looking beyond QCD, LEP measurements on the W -
mass~\cite{bib-straessner} and Tevatron
data~\cite{bib-cabrera,bib-zitoun} on the top mass continue to
improve the Standard Model parameters. Indeed, they continue to
support a light Higgs boson, although it may have gained some
weight \cite{bib-lepewwg}. The direct LEP limits constraining the
Standard Model Higgs mass from below and the Standard Model
radiative corrections from above, let the allowed mass range for
the Standard Model Higgs boson shrink to a rather small region.
However, to explore this range, or find out that the Higgs sector
is different, will still take some time. Recent Tevatron results
on neutral Higgs bosons in the decay $H\rightarrow W^+W^-$ or
doubly charged Higgs bosons do not reach a sensitivity to any
reasonable model \cite{bib-quizhongli}.

Whereas as yet measurements of W and Z bosons and top quarks are
rather limited, the accumulating luminosity at the Tevatron will
open new perspectives. Since the cross sections of W and Z
production are theoretically well understood \cite{bib-dorigo}
they may become a reference process for the luminosity at $pp$
collider, an important ingredient for precision physics at a
higher level. Top decays will become a very active field of
research at the Tevatron. They are sensitive to electroweak and
strong couplings and may reveal effects beyond the Standard Model.
First measurements with the Run I data have been performed
\cite{bib-zitoun}.

There is a significant sensitivity to new physics models from the
LEP and Tevatron data. The most popular extension of the Standard
Model is supersymmetry. However, no signal has been found in any
variant of supersymmetry~\cite{bib-braibant,bib-safonov}, be it
$R_p$ conserving or violating, gravity or gauge mediated. Within
the constrained MSSM with just five free parameters the results
can be translated into a limit on the lightest supersymmetric
particle, a dark matter candidate, to be heavier than 46 GeV.

Less well defined theoretically, with a larger range of possible
parameter sets are other models for Beyond the Standard Model
physics. Also for leptoquarks, a $Z'$, or other models, no signal
has been found~\cite{bib-goy,bib-hays}. A possible substructure,
as parametrized by contact interactions can be excluded, depending
on the detailed interaction, for scales up to 10 TeV.


\section{The future: near and far}

Our field is moving fast and new experimental opportunities are
opening up. Insights into many questions as yet unsettled
questions will soon be possible on several frontiers.

\begin{itemize}
\item RHIC is coming into gear,
\item PEPII and KEKB are improving their luminosities even
further to unprecedented levels,
\item Tevatron is on its way to select several fb$^{-1}$,
\item and HERA-II and CESR-c are about to start.
\end{itemize}

Extrapolating into the near future, we see that possible
highlights at Moriond 2006 will include precision measurements on
the top and W mass of 2 GeV, respectively a few 10s of MeV,
yielding a significant constraint on the Higgs mass. With some
luck, the Tevatron may even conclude on the Higgs mass with direct
measurements.

In 2007 a new era in High Energy physics will begin with the start
of data taking at the LHC. This heralds the beginning of a rather
complete exploration of the TeV scale. As was discussed in
\cite{bib-bruno} with an even modest start-up, LHC will dwarf all
existing data sets. With 1 fb$^{-1}$ (i.e. 3$\% $ of the perceived
luminosity at the first LHC year), for example, some 7 million
$W\rightarrow e,\mu \nu$, 80,000 $t\bar{t}$ pairs decaying
semileptonically will be produced. Once a few 10s of fb$^{-1}$
have been collected, a Standard Model Higgs could be observed
whatever its mass is~\cite{bib-mazumbdar}.

And beyond? The transcontinental consensus sees a linear $e^+e^-$
collider as the next project after LHC start - up. This is
underlined in documents from ACFA, HEPAP and ECFA. Launching this
truly world-wide accelerator will still require some way to go.
However, a first step is made. Recently, for the first time a
government has officially given support to this project. The
German government decided firstly to contribute several hundreds
of millions Euros to a Free Electron Laser built on the technology
developed for the Superconducting TESLA option. In building this
facility a significant insight into the industrial production of
SC cavities will be gained. Moreover it states that 'DESY will
continue its research work on TESLA in the existing international
framework, to facilitate German participation in a future global
project'. It is now our task to convince our respective
governments of the value of this exciting next project.

\section*{Acknowledgments}
The Rencontre de Moriond 2003 was again an inspiring meeting, for
which I would like to particularly thank Jean Tranh Tanh Van, Yuri
Dokshitzer and Boaz Klima. In preparing this talk I profited from
numerous discussions and help from a lot of people. To name a few,
I am grateful to John Butterworth, Susanna Carbrera, Peter Jacobs,
Katharina Klimek, George Lafferty, Gavin Salam and Bernd Surrow.
Etienne Auge and the whole secretariat gave me a a lot of support.
Finally I enjoyed technical assistance in preparing this write-up
from Andreas Kootz and Uwe M\"uller. Special thanks go to my
fellow theoretical summary speaker Lynne Orr for illuminating
discussions.

\section*{References}

If no explicit reference to a paper is given, they refer to
presentations at the conference.

\end{document}